 \documentclass[prl,10pt,twocolumn,showpacs,showkeys,preprintnumbers,amsmath,amssymb,superscriptaddress,aps]{revtex4-2}

\usepackage{graphicx}
\usepackage{dcolumn}
\usepackage{bm}
\usepackage{amsthm}
\usepackage{amsmath}
\usepackage{amssymb}
\usepackage{xcolor}

\begin{document}

\title{Generic role of the Dzyaloshinskii-Moriya interaction in nanocrystalline ferromagnets}

\author{Sergey~Erokhin}\email{s.erokhin@general-numerics-rl.de}
\affiliation{General Numerics Research Lab, Moritz-von-Rohr-Stra{\ss}e~1A, D-07745 Jena, Germany}
\author{Dmitry~Berkov}
\affiliation{General Numerics Research Lab, Moritz-von-Rohr-Stra{\ss}e~1A, D-07745 Jena, Germany}
\author{Andreas~Michels}
\affiliation{Department of Physics and Materials Science, University of Luxembourg, 162A~Avenue de la Faiencerie, L-1511 Luxembourg, Grand Duchy of Luxembourg}

\keywords{Dzyaloshinskii-Moriya interaction, polarized neutron scattering, micromagnetics, nanocrystalline terbium}

\begin{abstract}
Motivated by recent experimental polarized neutron results, we present a numerical micromagnetic study of the interfacial (intergrain) Dzyaloshinskii-Moriya interaction (DMI) in nanocrystalline terbium. We demonstrate that the DMI-induced spin misalignment between adjacent nanograins is the reason for the formation of the asymmetric positive-negative pattern seen in polarized neutron scattering experiments. Analysis of the remagnetization process suggests the generic impact of the DMI on the macroscopic magnetic parameters of polycrystalline defect-rich materials.
\end{abstract}

\maketitle

Originating from the relativistic spin-orbit coupling, the Dzyaloshinskii-Moriya interaction (DMI) is an antisymmetric contribution to the exchange energy between spins and plays a crucial role in a large variety of magnetic materials~\cite{dzyaloshinsky1958,moriya60}. In particular, in systems with a noncentrosymmetric crystal structure (lack of inversion symmetry), the DMI is essential for the formation of complex magnetization configurations, e.g., vortices, spin spirals, and skyrmions~\cite{bogdanov89,bogdanov94,pflei2009,yu2010,tokura2010,Nagaosa2013,rmp2019}. However, the noncentrosymmetric nature of the underlying crystal lattice is not the only mechanism by which antisymmetric exchange interactions are generated. In fact, it was predicted by Arrott~\cite{Arrott1963} that the DMI may be present in the vicinity of lattice defects of a crystal, where local breaking of structural inversion symmetry may take place. Besides, several other mechanisms such as the spin-orbit scattering of conduction electrons by nonmagnetic impurities in spin-glass alloys~\cite{fert1980}, the presence of structural inhomogeneities with an asymmetric distribution of the chemical composition in amorphous ferrimagnets~\cite{ono2019}, applied strain gradients~\cite{faehnle2010,beyerlein2018}, or the inversion-symmetry breaking at the interfaces of thin films~\cite{Boulle2016} are responsible for the appearance of antisymmetric exchange and the concomitant complex spin textures.

From the foregoing it becomes clear that a very important and broad class of materials which might exhibit DMI are {\it polycrystalline magnets}. One of the most prominent microstructural defects in polycrystalline materials are grain boundaries, which may be seen as two-dimensional interfaces separating crystallites of different crystallographic orientation. Consequently, DMI-induced changes in the magnetization configuration of such materials might be expected, in particular, in nanocrystalline magnets, which are polycrystalline materials with an average grain size $D_{\rm cr}$ of a few nanometers. Since the volume fraction of grain boundaries scales as $D_{\rm cr}^{-1}$, the effect should be largest for the smallest grain size. Indeed, analytical calculations performed in the framework of the continuum theory of micromagnetics have shown that the DMI should qualitatively affect the magnetization distribution and should manifest as an asymmetry of the polarized magnetic neutron scattering cross section~\cite{butenko2013,Michels2016PRB}. Recent neutron measurements on nanocrystalline terbium (Tb) have confirmed this important prediction~\cite{Michels2019PRB}.

\begin{figure*}[tb!]
\centering            
\resizebox{1.0\textwidth}{!}{\includegraphics{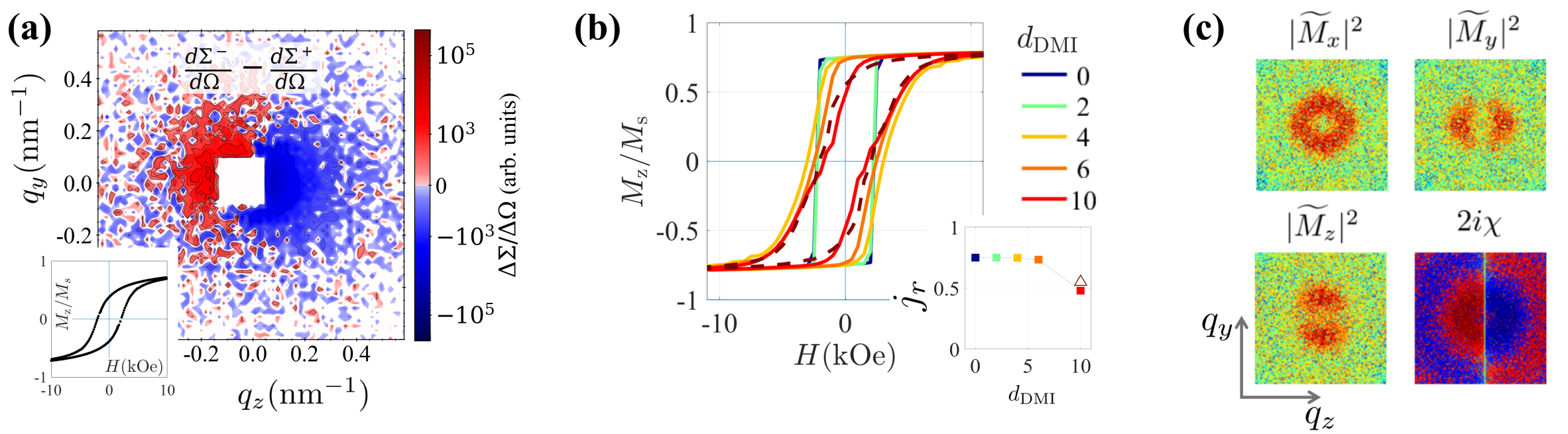}}
\caption{(a)~Experimental polarized neutron data of nanocrystalline Tb at $T = 100 \, \mathrm{K}$ and at an applied (horizontal) magnetic field of $B = 5 \, \mathrm{T}$~\cite{Michels2019PRB}. Shown is the difference between flipper-on ($d\Sigma^-/d\Omega$) and flipper-off ($d\Sigma^+/d\Omega$) small-angle neutron scattering cross sections. Experimental magnetization data are presented in the lower left corner. (b)~Simulated hysteresis loops for a system composed of $20 \, \mathrm{nm}$-sized Tb crystallites and for different DMI coefficient values. Solid lines---for a system with all interparticle coefficients $D^{\rm mes}_{\rm SW} > 0$. Reduced DMI constants shown in the legend are defined as $d_{\rm DMI} = D^{\rm mes}_{\rm SW} / D_0$, where $D_0 = 1.6 \times 10^{-12} \, \rm{erg}$ ($1 \, \rm{erg} = 10^{-7} \, \rm{J}$) [compare to Eq.~(\ref{eq:dmi_coef_SW}) in the Appendix]. Dashed loop corresponds to a system in which the DMI coefficients randomly change sign for different particle pairs ($D^{\rm mes}_{\rm SW} = \pm 1.6 \times 10^{-11}\, \rm{erg}$). (inset) Dependence of the reduced remanence $j_r = M_z(H=0)/M_s$ on the DMI value: squares---systems with $D^{\rm mes}_{\rm SW} > 0$; open triangle---a system with $D^{\rm mes}_{\rm SW} = \pm 1.6 \times 10^{-11}\, \rm{erg}$. (c)~Micromagnetic simulation results for the Fourier transforms of the magnetization components and for the resulting chiral function $2i \chi$ [Eq.~(\ref{eq:chiral_function})] for a system composed of $20 \, \mathrm{nm}$-sized crystallites with all positive DMI coefficients $D^{\rm mes}_{\rm SW} = + 1.6 \times 10^{-11} \, \mathrm{erg}$. Pixels in the corners of the images have $q = 0.42 \, \rm{nm^{-1}}$ (logarithmic color scales).}
\label{fig1}
\end{figure*}

The signature of defect-induced DMI is an asymmetric (positive-negative) pattern in the difference between polarized spin-up and spin-down neutron scattering cross sections [see Fig.~\ref{fig1}(a)]. This difference signal corresponds to chiral-type magnetization structures appearing in polycrystalline materials, even with a centrosymmetric crystal structure. In such crystals, the symmetry breaking might appear in the vicinity of the grain interfaces, resulting in a DMI between different grains. We note that the interlayer DMI effect has already been studied theoretically~\cite{Vedmedenko2019PRL}, and recently this idea has received an experimental validation~\cite{Avci2021}. As mentioned already above, considering that polycrystalline bulk ferromagnets represent a very broad and highly important class of materials (e.g., permanent magnets, magnetic steels, nanocomposites), micromagnetic simulations of the magnetization structures of such materials which include the DMI---the interaction that could lead to an asymmetry in the polarized neutron scattering cross section---are highly desirable to further understand this {\it generic} phenomenon.

The present study is organized as follows:~First, using a Stoner-Wohlfarth (SW) model, we summarize the modeling results obtained without the DMI. These simulations serve the purpose to understand the very basic behavior of nanocrystalline Tb, which possesses a relatively complex magnetic anisotropy energy. Second, we incorporate an interfacial DMI between the individual SW crystallites into the model, because it is in this system (without an internal magnetization structure of each individual crystallite) where we expect to obtain the strongest effect of the DMI interaction on the SANS pattern. At the final stage, we discuss the results of full-scale simulations of a three-dimensional system with large crystallites taking additionally into account the isotropic exchange and the magnetodipolar energies.

In all of the above steps, we have computed the three-dimensional magnetization vector field and the concomitant magnetic neutron scattering signal (the so-called chiral function). Details about the micromagnetic simulation methodology, in particular about the microstructure generation, the implementation of the various energy contributions, and the computation of the magnetic neutron scattering cross section, can be found in Refs.~\cite{michels2014jmmm,rmp2019}.


{\it Modeling without DMI}. First, to understand the basic behavior of Tb, we have performed micromagnetic simulations without the DMI. For this purpose, we have neglected the intergrain exchange interaction, because the experimental procedure used to obtain the nanocrystalline material under study (inert-gas condensation) leads to the formation of separated nanosized crystallites already in the gas phase, which are then pressed together to obtain a consolidated powder pellet~\cite{Michels2019PRB}. The resulting sample is expected to have highly disturbed intergrain boundaries, which should prevent the direct (symmetric) exchange coupling between the constituting grains. Furthermore, we have neglected the magnetodipolar interparticle interaction in this SW model, because the corresponding field is small compared to the large out-of-plane anisotropy field of Tb (a few tens of $\mathrm{T}$).

Single-domain behavior of Tb grains during the whole magnetization reversal process was observed for crystallites with a diameter below $D_{\rm cr} \lesssim 30 \, {\rm nm}$. Larger grains exhibit considerable deviations from the single-domain state. Such a relatively small critical single-domain size is due to the high magnetization of Tb (see Appendix), which leads to a very large demagnetizing field inside a grain in its homogeneous magnetization state, thus, favoring a transition to an inhomogeneous spin configuration. These findings allow us to apply a SW-like model (i.e., to use macrospins to represent the Tb crystallites) to a system of particles with a size smaller than $30 \, {\rm nm}$.



{\it Modelling including DMI}. To take into account the intergrain DMI in mesoscopic micromagnetic simulations, $E_{\rm DM}^{\rm mes}$, we have introduced the following contribution to the total micromagnetic energy, which is of the same functional form as the DMI energy between {\it atomic} magnetic moments $E_{\rm DM}^{\rm at}$, namely:
\begin{align}
\label{eq:dmi_atomic} 
E_{\rm DM}^{\rm at} &= \mathbf{D}^{\rm at} \cdot (\mathbf{S}_i \times \mathbf{S}_j) , \\    
\label{eq:dmi_mm} 
E_{\rm DM}^{\rm mes} &= \mathbf{D}^{\rm mes} \cdot (\mathbf{m}_i \times \mathbf{m}_j) ,
\end{align}
where the effective micromagnetic DMI vector $\mathbf{D}^{\rm mes}$ is parallel to the vector connecting the $i$-th and the $j$-th magnetic moment. We note already here that establishing a relation between the atomic and mesoscopic DMI constants (magnitudes of the DMI~vectors) is a nontrivial task requiring special attention (see Appendix).

Simulations of magnetization reversal in an ensemble of spherical single-domain crystallites with $D_{\rm cr} = 20 \, \mathrm{nm}$ including the DMI have also been performed using the SW-like model. Interparticle exchange and magnetodipolar interactions were neglected in this case for the reasons described above. Periodic boundary conditions were used. The total number of crystallites in our simulations is $N_{\rm cr} = 12500$, and the effective mesoscopic DMI parameter for this system was varied from zero to $|{\bf{D}}^{\rm mes}_{\rm SW}| = 1.6 \times 10^{-11} {\rm erg}$. When not stated otherwise, DMI coefficients for all particle pairs are positive. Note that the DMI values given here correspond to the {\it total} DMI energy between the $20 \, \mathrm{nm}$ large crystallites, and thus cannot be compared directly to the (much smaller) {\it interatomic} DMI coefficients.

Hysteresis loops are presented in Fig.~\ref{fig1}(b) and demonstrate that the magnetization reversal process substantially depends on the value of the effective DMI coefficient. A significant reduction of the reduced remanence $j_r$ with increasing DMI shown in the inset---from $j_r = 0.75$ for $D^{\rm mes}_{\rm SW} = 0.64 \times 10^{-11} \, \mathrm{erg}$ to $j_r = 0.48$ for $D^{\rm mes}_{\rm SW} = 1.6 \times 10^{-11} \, \mathrm{erg}$---can be explained by the tendency of the DMI to form spiral structures, which are inherently inhomogeneous and thus should reduce the remanence. Note that the remanence for a system without DMI is much larger than $j_r = 0.5$, because the anisotropy symmetry of the Tb grains is very much different from the simple uniaxial anisotropy assumed in the standard SW~model.

For comparison with the experimental polarized neutron data in Fig.~\ref{fig1}(a), we have computed the chiral function $\chi(\mathbf{q})$, which for the scattering geometry where the externally applied magnetic field $\mathbf{B}$ is perpendicular to the neutron beam can be expressed as~\cite{maleyev2002}:
\begin{equation}
\label{eq:chiral_function} 
\begin{split}
\chi(\mathbf{q}) = (\widetilde{M}_x \widetilde{M}_y^{\ast} - \widetilde{M}_x^{\ast} \widetilde{M}_y)\cos^2\theta \\ - (\widetilde{M}_x \widetilde{M}_z^{\ast} - \widetilde{M}_x^{\ast} \widetilde{M}_z)\sin\theta \cos\theta ,
\end{split}
\end{equation}
where the $\widetilde{M}_{x,y,z}(\mathbf{q})$ represent the Fourier transforms of the Cartesian magnetization components $M_{x,y,z}(\mathbf{r})$, and the asterisk ``$*$'' marks the complex-conjugated quantity. The (real-valued) quantity $2i \chi$, evaluated in the plane of the two-dimensional detector (corresponding to $q_x = 0$), can be directly compared to the experimental neutron data shown in Fig.~\ref{fig1}(a). We also emphasize that polarized neutron scattering is one of the few methods that is able to directly measure net chirality in a magnetic system. The squared amplitudes of the numerically-computed magnetization Fourier components along with the chiral function are presented in Fig.~\ref{fig1}(c). Note that here all DMI coefficients are positive. Comparison of the computed patterns for $2i \chi$ with the experimental data demonstrates a good qualitative agreement. Therefore, we conclude that the presence of DMI with either positive or negative sign (this is a matter of sign convention) results in magnetization configurations that give rise to the experimentally observed neutron patterns.

In order to further support the last statement, additional modeling has been carried out for a system where one half of the neighboring crystallite pairs are coupled via a positive DMI coefficient, while the other half are coupled via a negative one. The hysteresis loop for this case is presented in Fig.~\ref{fig1}(b) as the dashed line. While the difference between simulated loops for the cases when all the $D^{\rm mes}_{\rm SW} > 0$ and all the $D^{\rm mes}_{\rm SW}$ are randomly changing their sign is minor, the polarized neutron patterns for these two situations are significantly different: a system with equal fractions of positive and negative DMI coefficients demonstrates the complete lack of the chiral function. With all-positive DMI coefficients we artificially model the population asymmetry of the left-right helices experimentally observed in systems lacking inversion symmetry on the interfaces~\cite{Fedorov1997,lott08}.

\begin{figure*}[htb]
\centering
\resizebox{2.0\columnwidth}{!}{\includegraphics{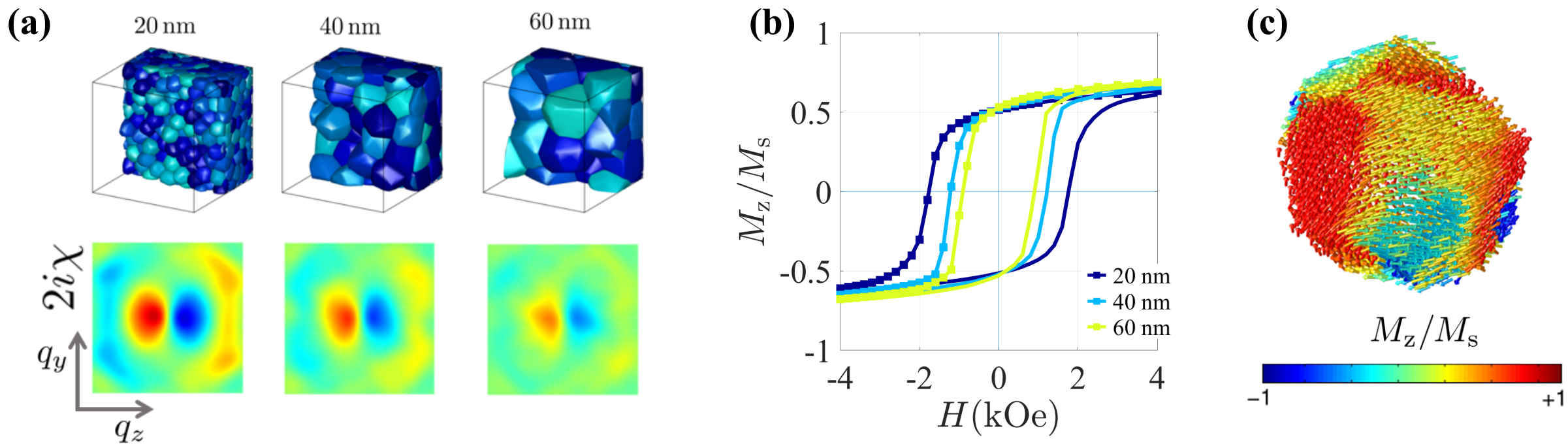}}
\caption{(a)~Examples of generated structures used in the micromagnetic modeling of nanocrystalline Tb with crystallize sizes of $20$, $40$, and $60 \, \mathrm{nm}$ along with the corresponding chiral functions $2i \chi$ at remanence (smoothed for better visibility, pixels in the corners have $q = 2.8 \, \rm{nm^{-1}}$, logarithmic color scale). (b)~Calculated hysteresis loops of nanocrystalline Tb for various grain sizes (averages over 8 geometrical realizations are presented). For this set of calculations, the DMI coefficient between neighboring mesh elements belonging to different grains is $D^{\rm mes}_{\rm mm} = 3.6 \times 10^{-13} \, {\rm erg} = 10  D^{\rm mes}_0$ [compare to Eq.~(\ref{eq:dmi_coef_MM}) in the Appendix]. (c)~Example for a magnetization distribution of a $60 \, \mathrm{nm}$-sized Tb grain at remanence. The depicted grain is a part of a system composed of many crystallites [see the upper row of subpanel~(a)].}
\label{fig2}
\end{figure*}


{\it Simulations of systems with large grains}. As explained above, larger crystallites may possess a complicated internal magnetization structure, so that simulations where each crystallite is adequately discretized in order to resolve this magnetization distribution are clearly necessary. In this case, we have used a cubical simulation box with a side length of $170 \, \mathrm{nm}$, subdivided into $\sim 2 \times 10^5$ mesh elements. Depending on their size, the total number of crystallites varies from $\sim 700$ for a $20 \, \mathrm{nm}$~crystallite sample to $26$ for a $60 \, \mathrm{nm}$~system. Examples of generated microstructures of polycrystalline Tb with various crystallite sizes are shown in the upper row of Fig.~\ref{fig2}(a).

In these simulations, we have used an exchange-stiffness constant of $A_{\rm ex} = 0.38 \times 10^{-6} \, \mathrm{erg/cm}$ between mesh elements belonging to the \textit{same} crystallite. This $A_{\rm ex}$~value was obtained by an additional procedure of mapping the atomistic magnetic parameters of Tb on a simple cubic lattice (chosen for simplicity) and fitting the obtained (by micromagnetic simulations) Bloch-wall profile to the well-known analytical solution. This procedure is necessary since there is no reliable analytical transformation from atomistic to mesoscopic parameters. The atomistic exchange constant was estimated from the Curie temperature of Tb (see Appendix). The exchange coupling between \textit{different} crystallites was neglected for the reasons described above. On the other hand, the DMI is expected to play an important role via the formation of noncentrosymmetric spin structures at the interfaces between crystallites, in this way leading to a nonzero chiral function. Hence, the DMI energy term [Eq.~(\ref{eq:dmi_mm})] was added for the mesh-element pairs for which the elements $i$ and $j$ belong to different crystallites. The magnetodipolar interaction was also taken into account and periodic boundary conditions were used. In these full-scale micromagnetic simulations, we define the reference DMI coefficient between mesh elements as $D^{\rm mes}_0 =  3.6 \times 10^{-14} \, \rm{erg}$.

For every set of structural and magnetic parameters, $8$ different realizations of the crystalline microstructure were simulated. Hysteresis loops averaged over these configurations are shown in Fig.~\ref{fig2}(b). While the remanence values and the approach-to-saturation behavior are very similar for all samples, the coercivity is strongly dependent on the crystallite size, being two times larger for a system with $D_{\rm cr} = 20 \, \mathrm{nm}$ as compared to the case of $D_{\rm cr} = 60 \, \mathrm{nm}$. This tendency is qualitatively the same as for the loops obtained without DMI (data not shown). 

The chiral function $2i \chi$ at remanence [Fig.~\ref{fig2}(a)] substantially changes with the crystallite size, which is in stark contrast to the corresponding magnetization value (remanence) that is almost size-independent [Fig.~\ref{fig2}(b)]. This observation demonstrates that the amount of information available by SANS, which is able to reveal the details of the spin structure in the bulk and on a mesoscopic length scale, is much larger than the information provided by integral (averaging) methods such as magnetometry. Importantly, in the full-scale micromagnetic simulations we find the same type of chiral function as for the SW-like model [compare Fig.~\ref{fig1}]. When the average grain size increases, the asymmetry of the pattern becomes less pronounced, since the ratio of the number of magnetic moments located at grain boundaries and the number of moments inside a crystallite decreases, thus, decreasing the relative importance of the DMI.

\begin{figure*}[htb]
\centering
\resizebox{1.95\columnwidth}{!}{\includegraphics{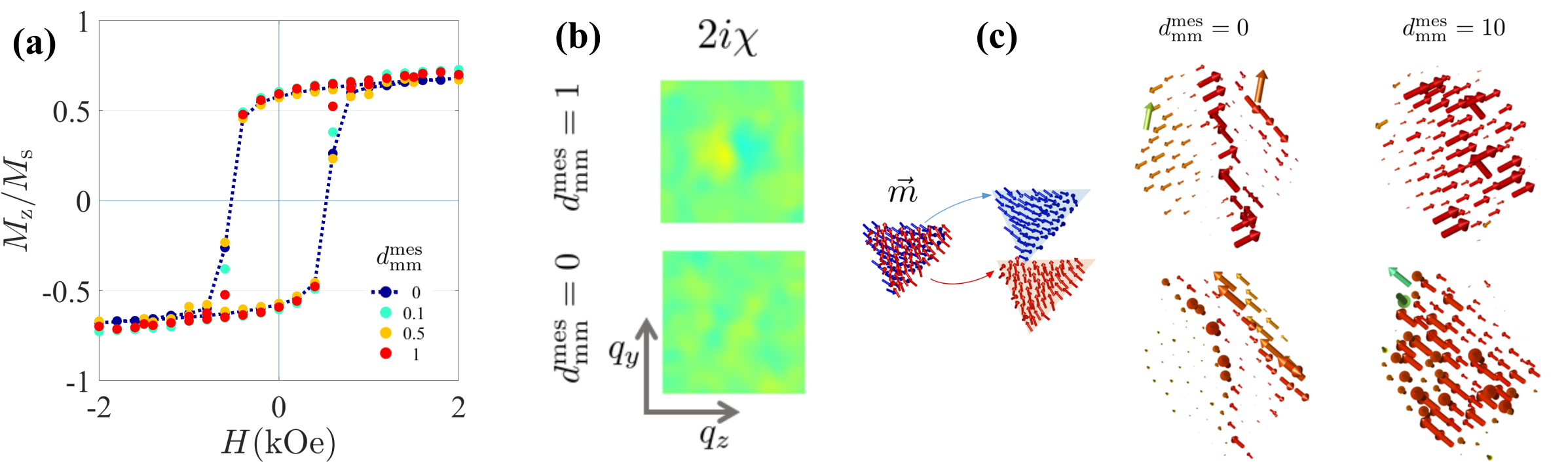}}
\caption{(a)~Simulated hysteresis loops (average over 8 geometrical realizations) of polycrystalline Tb consisting of $40 \, \mathrm{nm}$-sized grains for various normalized DMI coefficients $d^{\rm mes}_{\rm mm} = D^{\rm mes}_{\rm mm} / D^{\rm mes}_0$ (see inset). (b)~Corresponding chiral functions $2i \chi$ at remanence. Pixels in the corners have $q = 2.8 \, \rm{nm^{-1}}$ (logarithmic color scale). (c)~Magnetization distributions (shifted for better visibility) on the interface between two grains in a system composed of $40 \, {\rm nm}$ crystallites without and with DMI. The magnetization component (multiplied by a factor of 3) perpendicular to the anisotropy-axis direction nearest to the magnetization direction is shown for both cases.}
\label{fig3}
\end{figure*}

Additionally, we have studied the influence of the {\it magnitude} of the DMI value on the magnetization reversal and the corresponding SANS pattern (Fig.~\ref{fig3}). For this purpose, hysteresis loops of a system composed of $40 \, \mathrm{nm}$-sized grains have been simulated for a range of DMI coefficients that are an order of magnitude smaller than for the simulation results shown in Fig.~\ref{fig2}(b) ($D^{\rm mes}_{\rm mm} =  3.6 \times 10^{-14} \, {\rm erg} = D^{\rm mes}_0$). Due to this smaller value of the DMI constant, the remanence for the loops shown in Fig.~\ref{fig3}(a) is slightly larger than for those presented in Fig.~\ref{fig2}(b) (light-blue curve), because a decreasing DMI leads to the reduction of the spin misalignment at the intergrain interfaces. The main result of this set of simulations is the absence of a statistically-relevant difference between magnetization curves for DMI constants in the range $d^{\rm mes}_{\rm mm} = D^{\rm mes}_{\rm mm} / D^{\rm mes}_0 = 0$$-$$1.0$.

On the other hand, the chiral function $2i \chi$ shown in Fig.~\ref{fig3}(b) demonstrates a qualitative difference between systems with $d^{\rm mes}_{\rm mm} = 1$ and $d^{\rm mes}_{\rm mm} = 10$ [see Fig.~\ref{fig2}(a), $40 \, \mathrm{nm}$ grain size data] and without DMI ($d^{\rm mes}_{\rm mm} = 0$). Namely, the strong contrast of the positive-negative pattern of $2i \chi$ of the sample with the largest interfacial DMI decreases in the system with the reduced DMI and completely disappears without this interaction. Note that the corresponding pattern is still clearly visible for $d^{\rm mes}_{\rm mm} = 1$, where the hysteresis loop is identical to that of the system without DMI [Fig.~\ref{fig3}(a)]. This observation confirms once more the power of the SANS method to reveal fine features of the bulk magnetization structure.

Our simulations also allow to understand the effect of the DMI on the details of the magnetization configuration, in particular, on the correlation of magnetization states in neighboring grains. The magnetization distribution in the polycrystalline system under study when the interfacial DMI is included (and competes with the other interactions) is very complex, as it can be seen for the example displayed in Fig.~\ref{fig2}(c). To understand the changes in the spin structure due to the presence of the DMI, which here is an interface-mediated interaction, we have compared the orientations of magnetic moments at the interface between crystallites for identical systems without and with DMI. It turns out that these changes are not only quantitative, but also qualitative, as it is presented in Fig.~\ref{fig3}(c). Here, we show the influence of the DMI on the magnetization vector field at the interface between adjacent grains. In the system without DMI, there exists a kind of magnetic domain boundary separating two domains with magnetization orientations along two different directions of easy axes (we remind that there exist a six-fold anisotropy in the easy plane of Tb, corresponding to 3 easy axes in this plane). On the contrary, for the system with DMI, we do not observe any sharply-defined domain wall, but find strong deviations from the easy axes on the entire intergrain interface.


Summarizing, we have developed a micromagnetic model of nanocrystalline Tb with a strong easy-plane magnetic anisotropy and an additional sixth-fold anisotropy within this easy plane. It has been shown that the intergrain (interfacial) Dzyaloshinskii-Moriya interaction (DMI) which is incorporated into our approach results in an additional spin misalignment between adjacent nanocrystalline Tb grains. The magnetization reversal processes has been studied in detail both in frames of a Stoner-Wohlfarth-like model (for small grain sizes) and using full-scale micromagnetic simulations (for systems with larger crystallite sizes where the internal magnetization distribution is important). Both approaches have demonstrated that the DMI has a pronounced effect on the macroscopic magnetic parameters and that it is at the origin of the asymmetric positive-negative pattern of the polarized neutron scattering signal observed experimentally. These results underline the generic role of the DMI for the magnetism of defect-rich polycrystalline magnets.

\section*{Appendix: Materials parameters and relations between the various DMI coefficients}

Terbium crystallizes in a hexagonal closed-packed (HCP) structure with lattice constants of $a = 3.6055 \, \mbox{\normalfont{\AA}}$ and $c = 5.6966 \, \mbox{\normalfont{\AA}}$ ($c/a \cong 1.58$)~\cite{legvold80}. Its magnetism originates from the electrons in the partially filled $4f$~shell, which give rise to localized magnetic moments that couple via the long-range Ruderman-Kittel-Kasuya-Yosida (RKKY) interaction involving the conduction electrons. The Tb$^{3+}$~ion has a spin angular momentum of $S=3$ and an orbital angular momentum of $L=3$, which results in a highly anisotropic electronic charge cloud and in a concomitant complex magnetocrystalline anisotropy. Below $T_c = 220 \, \mathrm{K}$, the magnetic moments are confined by an extremely large magnetocrystalline anisotropy (of strength $K_1 = 6 \times 10^8 \, \mathrm{erg/cm^3}$) into the basal plane of the HCP lattice~\cite{chikazumi1997book}. Within the basal plane, there exists an additional six-fold anisotropy, with a corresponding (weaker) anisotropy constant of $K_6^6 = 0.6 \times 10^6 \, \mathrm{erg/cm^3}$. In the simulations, we used the following expression for the anisotropy-energy density of Tb:
\begin{eqnarray}
\label{eq:ani_function} 
\epsilon_A = K_1 \cos^2(\alpha) + K_6^6 \sin^2(\alpha) \cos(6\beta) ,
\end{eqnarray}
where $\alpha$ and $\beta$ are the angles of the magnetization with respect to the $c$ and $a$~axes of the HCP crystal.

The difference in magnitude between the two anisotropy constants and the polycrystalline nature of the sample may then qualitatively explain the hard-soft behavior of the experimentally obtained magnetization curve [see Fig.~\ref{fig1}(a)]. Due to the large $K_1$~value it is not possible to produce a significant tilting of magnetic moments out of the easy plane with available laboratory magnetic fields (say, fields smaller than $10 \, \mathrm{T}$). On the other hand, within the easy plane, a magnetic moment has to overcome only a relatively low energy barrier produced by a moderate anisotropy $K_6^6$ (compared to $K_1$), so that its orientation can be reversed within this easy plane already by a relatively small negative field. This scenario then effectively leads to a kind of hard-soft behavior of the experimental hysteresis loop. 

The experimental determination of the saturation magnetization of nanocrystalline Tb is challenging due to the extremely large anisotropy, so that for the micromagnetic simulations the saturation magnetization value at $100 \, \mathrm{K}$ of single crystalline Tb was taken, $M_s \cong 2354 \, \mathrm{kA/m}$~\cite{hegland63}. The exchange constant has been estimated from the well-known mean-field relation $J_{\mathrm{ex}} = 3 k_B T_c/(\epsilon z)$~\cite{garanin1992,evans2014}, where $k_B = 1.381 \times 10^{-16} \, \mathrm{erg/K}$, $z$ is the number of nearest neighbors, and $\epsilon$ is a correction factor of the order of unity arising from the consideration of spin waves. In this way, we find $J_{\mathrm{ex}} = 9.1 \times 10^{-15} \, \mathrm{erg}$.

The relation between the DMI coefficients used in our mesoscopic simulations ($D^{\rm mes}_{\rm SW}$ and $D^{\rm mes}_{\rm mm}$) and the corresponding atomistic values ($D^{\rm at}$) can be established by estimating the DMI energy $E_{\rm DM}^{\rm g-g}$ of the interface between two grains as:
\begin{equation}
\label{eq:dmi_energy_grains} 
E_{\rm DM}^{\rm g-g} = 
{\bf D}^{\rm mes} \cdot ({\bf m}_i \times {\bf m}_j) = 
N^{\rm g-g} \, {\bf D}^{\rm at} \cdot ({\bf S}_i \times {\bf S}_j) ,
\end{equation}
where ${\bf m}_i$ and ${\bf S}_i$ denote the unit vectors of, respectively, the magnetic moment of the $i$-th crystallite and of the spin of the $i$-th atom at the interface. This yields a simple relation between the average atomic DMI constant $D^{\rm at}$ and its mesoscopic counterpart $D^{\rm mes}$ using the number $N^{\rm g-g}$ of atoms at the interface:
\begin{equation}
\label{eq:dmi_atomic_coef} 
D^{\rm at} = \frac{D^{\rm mes}}{N^{\rm g-g}} =
 \frac{D^{\rm mes}}{n_{\rm surf} S^{\rm g-g}} .
\end{equation}
Here, the atomic surface density is defined as $n_{\rm surf} = n_{\rm at}/\bar{l}^2$, where $n_{\rm at} = 2$ is the number of atoms in the elementary HCP unit cell, and $\bar{l} \cong 4.30 \, \mbox{\normalfont{\AA}}$ is the corresponding average cell size for Tb. The average surface area of the intergrain interfaces, $S^{\rm g-g}$, can be computed from the average number of nearest neighboring grains, $\langle k_{\rm av} \rangle \cong 12.6$, as $S^{\rm g-g} = 4\pi r^2_{\rm g} / \langle k_{\rm av} \rangle$, where $r_{\rm g}$ denotes the grain's radius.

The conversion to the often employed DMI coefficient $D^{\rm sed}$, which has the unit of a {\it surface energy density} ($1 \, \rm{erg/cm^2} = 1 \, \rm{mJ/m^2}$), is obtained by a similar relation using the interface energy:
\begin{equation}
\label{eq:dmi_coef_SW} 
E_{\rm DM (SW)}^{\rm g-g} \equiv D^{\rm mes}_{\rm SW} = D^{\rm sed} S^{\rm g-g} .
\end{equation}
For the case of uniformly magnetized $20 \, \rm{nm}$-sized Stoner-Wohlfarth particles (and using $E_{\rm DM (SW)}^{\rm g-g} = 1.6 \times 10^{-12} \, \rm{erg}$ and $n_{\rm surf} \cong 1.08 \times 10^{15} \, \mathrm{cm}^{-2}$), we obtain $D^{\rm at} \cong 1.48 \times 10^{-15} \, {\rm erg}$ and $D^{\rm sed} \cong 1.60 \, {\rm erg/cm^2}$.

In the full-scale micromagnetic calculations (with discretized grains), we deal with
\begin{equation}
\label{eq:dmi_coef_MM} 
E_{\rm DM (mm)}^{\rm g-g} = N^{\rm g-g}_{\rm mesh} D^{\rm mes}_0 ,
\end{equation}
where the number of mesh elements per interface area for the $20 \, \rm{nm}$-sized grain is 
$N^{\rm g-g}_{\rm mesh} = S^{\rm g-g}/(\pi r^2_{\rm mesh}) \cong 14$ (using $2r_{\rm mesh} = 3 \, \rm{nm}$). In this case, for $D^{\rm mes}_0 = 3.6 \times 10^{-14} \, \rm{erg}$ and hence $E_{\rm DM (mm)}^{\rm g-g} = 5.0 \times 10^{-13} \, \rm{erg}$, the DMI coefficients corresponding to our micromagnetic simulations are $D^{\rm at} \cong 4.64 \times 10^{-16} \, {\rm erg}$ and $D^{\rm sed} \cong 0.50 \, {\rm erg/cm^2}$. In Ref.~\cite{Michels2019PRB}, the value for the DMI coefficient of nanocrystalline Tb has been found as: $D^{\rm mes}_{\rm Tb} = 0.45 \, \rm{mJ/m^2} = 0.45 \, {\rm erg/cm^2}$.


\begin{thebibliography}{29}%
\makeatletter
\providecommand \@ifxundefined [1]{%
 \@ifx{#1\undefined}
}%
\providecommand \@ifnum [1]{%
 \ifnum #1\expandafter \@firstoftwo
 \else \expandafter \@secondoftwo
 \fi
}%
\providecommand \@ifx [1]{%
 \ifx #1\expandafter \@firstoftwo
 \else \expandafter \@secondoftwo
 \fi
}%
\providecommand \natexlab [1]{#1}%
\providecommand \enquote  [1]{``#1''}%
\providecommand \bibnamefont  [1]{#1}%
\providecommand \bibfnamefont [1]{#1}%
\providecommand \citenamefont [1]{#1}%
\providecommand \href@noop [0]{\@secondoftwo}%
\providecommand \href [0]{\begingroup \@sanitize@url \@href}%
\providecommand \@href[1]{\@@startlink{#1}\@@href}%
\providecommand \@@href[1]{\endgroup#1\@@endlink}%
\providecommand \@sanitize@url [0]{\catcode `\\12\catcode `\$12\catcode
  `\&12\catcode `\#12\catcode `\^12\catcode `\_12\catcode `\%12\relax}%
\providecommand \@@startlink[1]{}%
\providecommand \@@endlink[0]{}%
\providecommand \url  [0]{\begingroup\@sanitize@url \@url }%
\providecommand \@url [1]{\endgroup\@href {#1}{\urlprefix }}%
\providecommand \urlprefix  [0]{URL }%
\providecommand \Eprint [0]{\href }%
\providecommand \doibase [0]{https://doi.org/}%
\providecommand \selectlanguage [0]{\@gobble}%
\providecommand \bibinfo  [0]{\@secondoftwo}%
\providecommand \bibfield  [0]{\@secondoftwo}%
\providecommand \translation [1]{[#1]}%
\providecommand \BibitemOpen [0]{}%
\providecommand \bibitemStop [0]{}%
\providecommand \bibitemNoStop [0]{.\EOS\space}%
\providecommand \EOS [0]{\spacefactor3000\relax}%
\providecommand \BibitemShut  [1]{\csname bibitem#1\endcsname}%
\let\auto@bib@innerbib\@empty
\bibitem [{\citenamefont {Dzyaloshinsky}(1958)}]{dzyaloshinsky1958}%
  \BibitemOpen
  \bibfield  {author} {\bibinfo {author} {\bibfnamefont {I.}~\bibnamefont
  {Dzyaloshinsky}},\ }\href@noop {} {\bibfield  {journal} {\bibinfo  {journal}
  {J. Phys. Chem. Solids}\ }\textbf {\bibinfo {volume} {4}},\ \bibinfo {pages}
  {241} (\bibinfo {year} {1958})}\BibitemShut {NoStop}%
\bibitem [{\citenamefont {Moriya}(1960)}]{moriya60}%
  \BibitemOpen
  \bibfield  {author} {\bibinfo {author} {\bibfnamefont {T.}~\bibnamefont
  {Moriya}},\ }\href@noop {} {\bibfield  {journal} {\bibinfo  {journal} {Phys.
  Rev.}\ }\textbf {\bibinfo {volume} {120}},\ \bibinfo {pages} {91} (\bibinfo
  {year} {1960})}\BibitemShut {NoStop}%
\bibitem [{\citenamefont {Bogdanov}\ and\ \citenamefont
  {Yablonski\u{\i}}(1989)}]{bogdanov89}%
  \BibitemOpen
  \bibfield  {author} {\bibinfo {author} {\bibfnamefont {A.~N.}\ \bibnamefont
  {Bogdanov}}\ and\ \bibinfo {author} {\bibfnamefont {D.~A.}\ \bibnamefont
  {Yablonski\u{\i}}},\ }\href@noop {} {\bibfield  {journal} {\bibinfo
  {journal} {Sov. Phys. JETP}\ }\textbf {\bibinfo {volume} {68}},\ \bibinfo
  {pages} {101} (\bibinfo {year} {1989})}\BibitemShut {NoStop}%
\bibitem [{\citenamefont {Bogdanov}\ and\ \citenamefont
  {Hubert}(1994)}]{bogdanov94}%
  \BibitemOpen
  \bibfield  {author} {\bibinfo {author} {\bibfnamefont {A.}~\bibnamefont
  {Bogdanov}}\ and\ \bibinfo {author} {\bibfnamefont {A.}~\bibnamefont
  {Hubert}},\ }\href@noop {} {\bibfield  {journal} {\bibinfo  {journal} {J.
  Magn. Magn. Mater.}\ }\textbf {\bibinfo {volume} {138}},\ \bibinfo {pages}
  {255} (\bibinfo {year} {1994})}\BibitemShut {NoStop}%
\bibitem [{\citenamefont {M\"uhlbauer}\ \emph {et~al.}(2009)\citenamefont
  {M\"uhlbauer}, \citenamefont {Binz}, \citenamefont {Jonietz}, \citenamefont
  {Pfleiderer}, \citenamefont {Rosch}, \citenamefont {Neubauer}, \citenamefont
  {Georgii},\ and\ \citenamefont {B\"oni}}]{pflei2009}%
  \BibitemOpen
  \bibfield  {author} {\bibinfo {author} {\bibfnamefont {S.}~\bibnamefont
  {M\"uhlbauer}}, \bibinfo {author} {\bibfnamefont {B.}~\bibnamefont {Binz}},
  \bibinfo {author} {\bibfnamefont {F.}~\bibnamefont {Jonietz}}, \bibinfo
  {author} {\bibfnamefont {C.}~\bibnamefont {Pfleiderer}}, \bibinfo {author}
  {\bibfnamefont {A.}~\bibnamefont {Rosch}}, \bibinfo {author} {\bibfnamefont
  {A.}~\bibnamefont {Neubauer}}, \bibinfo {author} {\bibfnamefont
  {R.}~\bibnamefont {Georgii}},\ and\ \bibinfo {author} {\bibfnamefont
  {P.}~\bibnamefont {B\"oni}},\ }\href@noop {} {\bibfield  {journal} {\bibinfo
  {journal} {Science}\ }\textbf {\bibinfo {volume} {323}},\ \bibinfo {pages}
  {915} (\bibinfo {year} {2009})}\BibitemShut {NoStop}%
\bibitem [{\citenamefont {Yu}\ \emph {et~al.}(2010)\citenamefont {Yu},
  \citenamefont {Onose}, \citenamefont {Kanazawa}, \citenamefont {Park},
  \citenamefont {Han}, \citenamefont {Matsui}, \citenamefont {Nagaosa},\ and\
  \citenamefont {Tokura}}]{yu2010}%
  \BibitemOpen
  \bibfield  {author} {\bibinfo {author} {\bibfnamefont {X.~Z.}\ \bibnamefont
  {Yu}}, \bibinfo {author} {\bibfnamefont {Y.}~\bibnamefont {Onose}}, \bibinfo
  {author} {\bibfnamefont {N.}~\bibnamefont {Kanazawa}}, \bibinfo {author}
  {\bibfnamefont {J.~H.}\ \bibnamefont {Park}}, \bibinfo {author}
  {\bibfnamefont {J.~H.}\ \bibnamefont {Han}}, \bibinfo {author} {\bibfnamefont
  {Y.}~\bibnamefont {Matsui}}, \bibinfo {author} {\bibfnamefont
  {N.}~\bibnamefont {Nagaosa}},\ and\ \bibinfo {author} {\bibfnamefont
  {Y.}~\bibnamefont {Tokura}},\ }\href@noop {} {\bibfield  {journal} {\bibinfo
  {journal} {Nature}\ }\textbf {\bibinfo {volume} {465}},\ \bibinfo {pages}
  {901} (\bibinfo {year} {2010})}\BibitemShut {NoStop}%
\bibitem [{\citenamefont {Tokura}\ and\ \citenamefont
  {Seki}(2010)}]{tokura2010}%
  \BibitemOpen
  \bibfield  {author} {\bibinfo {author} {\bibfnamefont {Y.}~\bibnamefont
  {Tokura}}\ and\ \bibinfo {author} {\bibfnamefont {S.}~\bibnamefont {Seki}},\
  }\href@noop {} {\bibfield  {journal} {\bibinfo  {journal} {Adv. Mater.}\
  }\textbf {\bibinfo {volume} {22}},\ \bibinfo {pages} {1554} (\bibinfo {year}
  {2010})}\BibitemShut {NoStop}%
\bibitem [{\citenamefont {Nagaosa}\ and\ \citenamefont
  {Tokura}(2013)}]{Nagaosa2013}%
  \BibitemOpen
  \bibfield  {author} {\bibinfo {author} {\bibfnamefont {N.}~\bibnamefont
  {Nagaosa}}\ and\ \bibinfo {author} {\bibfnamefont {Y.}~\bibnamefont
  {Tokura}},\ }\href@noop {} {\bibfield  {journal} {\bibinfo  {journal} {Nat.
  Nanotech.}\ }\textbf {\bibinfo {volume} {8}},\ \bibinfo {pages} {899}
  (\bibinfo {year} {2013})}\BibitemShut {NoStop}%
\bibitem [{\citenamefont {M\"uhlbauer}\ \emph {et~al.}(2019)\citenamefont
  {M\"uhlbauer}, \citenamefont {Honecker}, \citenamefont {P\'{e}rigo},
  \citenamefont {Bergner}, \citenamefont {Disch}, \citenamefont {Heinemann},
  \citenamefont {Erokhin}, \citenamefont {Berkov}, \citenamefont {Leighton},
  \citenamefont {Eskildsen},\ and\ \citenamefont {Michels}}]{rmp2019}%
  \BibitemOpen
  \bibfield  {author} {\bibinfo {author} {\bibfnamefont {S.}~\bibnamefont
  {M\"uhlbauer}}, \bibinfo {author} {\bibfnamefont {D.}~\bibnamefont
  {Honecker}}, \bibinfo {author} {\bibfnamefont {E.~A.}\ \bibnamefont
  {P\'{e}rigo}}, \bibinfo {author} {\bibfnamefont {F.}~\bibnamefont {Bergner}},
  \bibinfo {author} {\bibfnamefont {S.}~\bibnamefont {Disch}}, \bibinfo
  {author} {\bibfnamefont {A.}~\bibnamefont {Heinemann}}, \bibinfo {author}
  {\bibfnamefont {S.}~\bibnamefont {Erokhin}}, \bibinfo {author} {\bibfnamefont
  {D.}~\bibnamefont {Berkov}}, \bibinfo {author} {\bibfnamefont
  {C.}~\bibnamefont {Leighton}}, \bibinfo {author} {\bibfnamefont {M.~R.}\
  \bibnamefont {Eskildsen}},\ and\ \bibinfo {author} {\bibfnamefont
  {A.}~\bibnamefont {Michels}},\ }\href@noop {} {\bibfield  {journal} {\bibinfo
   {journal} {Rev. Mod. Phys.}\ }\textbf {\bibinfo {volume} {91}},\ \bibinfo
  {pages} {015004} (\bibinfo {year} {2019})}\BibitemShut {NoStop}%
\bibitem [{\citenamefont {Arrott}(1963)}]{Arrott1963}%
  \BibitemOpen
  \bibfield  {author} {\bibinfo {author} {\bibfnamefont {A.}~\bibnamefont
  {Arrott}},\ }\href {https://doi.org/10.1063/1.1729390} {\bibfield  {journal}
  {\bibinfo  {journal} {J. Appl. Phys.}\ }\textbf {\bibinfo {volume} {34}},\
  \bibinfo {pages} {1108} (\bibinfo {year} {1963})}\BibitemShut {NoStop}%
\bibitem [{\citenamefont {Fert}\ and\ \citenamefont {Levy}(1980)}]{fert1980}%
  \BibitemOpen
  \bibfield  {author} {\bibinfo {author} {\bibfnamefont {A.}~\bibnamefont
  {Fert}}\ and\ \bibinfo {author} {\bibfnamefont {P.~M.}\ \bibnamefont
  {Levy}},\ }\href@noop {} {\bibfield  {journal} {\bibinfo  {journal} {Phys.
  Rev. Lett.}\ }\textbf {\bibinfo {volume} {44}},\ \bibinfo {pages} {1538}
  (\bibinfo {year} {1980})}\BibitemShut {NoStop}%
\bibitem [{\citenamefont {Kim}\ \emph {et~al.}(2019)\citenamefont {Kim},
  \citenamefont {Haruta}, \citenamefont {Ko}, \citenamefont {Go}, \citenamefont
  {Park}, \citenamefont {Nishimura}, \citenamefont {Kim}, \citenamefont
  {Okuno}, \citenamefont {Hirata}, \citenamefont {Futakawa}, \citenamefont
  {Yoshikawa}, \citenamefont {Ham}, \citenamefont {Kim}, \citenamefont
  {Kurata}, \citenamefont {Tsukamoto}, \citenamefont {Shiota}, \citenamefont
  {Moriyama}, \citenamefont {Choe}, \citenamefont {Lee},\ and\ \citenamefont
  {Ono}}]{ono2019}%
  \BibitemOpen
  \bibfield  {author} {\bibinfo {author} {\bibfnamefont {D.-H.}\ \bibnamefont
  {Kim}}, \bibinfo {author} {\bibfnamefont {M.}~\bibnamefont {Haruta}},
  \bibinfo {author} {\bibfnamefont {H.-W.}\ \bibnamefont {Ko}}, \bibinfo
  {author} {\bibfnamefont {G.}~\bibnamefont {Go}}, \bibinfo {author}
  {\bibfnamefont {H.-J.}\ \bibnamefont {Park}}, \bibinfo {author}
  {\bibfnamefont {T.}~\bibnamefont {Nishimura}}, \bibinfo {author}
  {\bibfnamefont {D.-Y.}\ \bibnamefont {Kim}}, \bibinfo {author} {\bibfnamefont
  {T.}~\bibnamefont {Okuno}}, \bibinfo {author} {\bibfnamefont
  {Y.}~\bibnamefont {Hirata}}, \bibinfo {author} {\bibfnamefont
  {Y.}~\bibnamefont {Futakawa}}, \bibinfo {author} {\bibfnamefont
  {H.}~\bibnamefont {Yoshikawa}}, \bibinfo {author} {\bibfnamefont
  {W.}~\bibnamefont {Ham}}, \bibinfo {author} {\bibfnamefont {S.}~\bibnamefont
  {Kim}}, \bibinfo {author} {\bibfnamefont {H.}~\bibnamefont {Kurata}},
  \bibinfo {author} {\bibfnamefont {A.}~\bibnamefont {Tsukamoto}}, \bibinfo
  {author} {\bibfnamefont {Y.}~\bibnamefont {Shiota}}, \bibinfo {author}
  {\bibfnamefont {T.}~\bibnamefont {Moriyama}}, \bibinfo {author}
  {\bibfnamefont {S.-B.}\ \bibnamefont {Choe}}, \bibinfo {author}
  {\bibfnamefont {K.-J.}\ \bibnamefont {Lee}},\ and\ \bibinfo {author}
  {\bibfnamefont {T.}~\bibnamefont {Ono}},\ }\href@noop {} {\bibfield
  {journal} {\bibinfo  {journal} {Nat. Mater.}\ }\textbf {\bibinfo {volume}
  {18}},\ \bibinfo {pages} {685} (\bibinfo {year} {2019})}\BibitemShut
  {NoStop}%
\bibitem [{\citenamefont {Beck}\ and\ \citenamefont
  {F\"ahnle}(2010)}]{faehnle2010}%
  \BibitemOpen
  \bibfield  {author} {\bibinfo {author} {\bibfnamefont {P.}~\bibnamefont
  {Beck}}\ and\ \bibinfo {author} {\bibfnamefont {M.}~\bibnamefont
  {F\"ahnle}},\ }\href@noop {} {\bibfield  {journal} {\bibinfo  {journal} {J.
  Magn. Magn. Mater.}\ }\textbf {\bibinfo {volume} {322}},\ \bibinfo {pages}
  {3701} (\bibinfo {year} {2010})}\BibitemShut {NoStop}%
\bibitem [{\citenamefont {Kitchaev}\ \emph {et~al.}(2018)\citenamefont
  {Kitchaev}, \citenamefont {Beyerlein},\ and\ \citenamefont {Van~der
  Ven}}]{beyerlein2018}%
  \BibitemOpen
  \bibfield  {author} {\bibinfo {author} {\bibfnamefont {D.~A.}\ \bibnamefont
  {Kitchaev}}, \bibinfo {author} {\bibfnamefont {I.~J.}\ \bibnamefont
  {Beyerlein}},\ and\ \bibinfo {author} {\bibfnamefont {A.}~\bibnamefont
  {Van~der Ven}},\ }\href@noop {} {\bibfield  {journal} {\bibinfo  {journal}
  {Phys. Rev. B}\ }\textbf {\bibinfo {volume} {98}},\ \bibinfo {pages} {214414}
  (\bibinfo {year} {2018})}\BibitemShut {NoStop}%
\bibitem [{\citenamefont {Boulle}\ \emph {et~al.}(2016)\citenamefont {Boulle},
  \citenamefont {Vogel}, \citenamefont {Yang}, \citenamefont {Pizzini},
  \citenamefont {de~Souza~Chaves}, \citenamefont {Locatelli}, \citenamefont
  {Mente\c{s}}, \citenamefont {Sala}, \citenamefont {Buda-Prejbeanu},
  \citenamefont {Klein}, \citenamefont {Belmeguenai}, \citenamefont
  {Roussign\'e}, \citenamefont {Stashkevich}, \citenamefont {Ch\'erif},
  \citenamefont {Aballe}, \citenamefont {Foerster}, \citenamefont {Chshiev},
  \citenamefont {Auffret}, \citenamefont {Miron},\ and\ \citenamefont
  {Gaudin}}]{Boulle2016}%
  \BibitemOpen
  \bibfield  {author} {\bibinfo {author} {\bibfnamefont {O.}~\bibnamefont
  {Boulle}}, \bibinfo {author} {\bibfnamefont {J.}~\bibnamefont {Vogel}},
  \bibinfo {author} {\bibfnamefont {H.}~\bibnamefont {Yang}}, \bibinfo {author}
  {\bibfnamefont {S.}~\bibnamefont {Pizzini}}, \bibinfo {author} {\bibfnamefont
  {D.}~\bibnamefont {de~Souza~Chaves}}, \bibinfo {author} {\bibfnamefont
  {A.}~\bibnamefont {Locatelli}}, \bibinfo {author} {\bibfnamefont {T.~O.}\
  \bibnamefont {Mente\c{s}}}, \bibinfo {author} {\bibfnamefont
  {A.}~\bibnamefont {Sala}}, \bibinfo {author} {\bibfnamefont {L.~D.}\
  \bibnamefont {Buda-Prejbeanu}}, \bibinfo {author} {\bibfnamefont
  {O.}~\bibnamefont {Klein}}, \bibinfo {author} {\bibfnamefont
  {M.}~\bibnamefont {Belmeguenai}}, \bibinfo {author} {\bibfnamefont
  {Y.}~\bibnamefont {Roussign\'e}}, \bibinfo {author} {\bibfnamefont
  {A.}~\bibnamefont {Stashkevich}}, \bibinfo {author} {\bibfnamefont {S.~M.}\
  \bibnamefont {Ch\'erif}}, \bibinfo {author} {\bibfnamefont {L.}~\bibnamefont
  {Aballe}}, \bibinfo {author} {\bibfnamefont {M.}~\bibnamefont {Foerster}},
  \bibinfo {author} {\bibfnamefont {M.}~\bibnamefont {Chshiev}}, \bibinfo
  {author} {\bibfnamefont {S.}~\bibnamefont {Auffret}}, \bibinfo {author}
  {\bibfnamefont {I.~M.}\ \bibnamefont {Miron}},\ and\ \bibinfo {author}
  {\bibfnamefont {G.}~\bibnamefont {Gaudin}},\ }\href
  {https://doi.org/10.1038/nnano.2015.315} {\bibfield  {journal} {\bibinfo
  {journal} {Nature Nanotech.}\ }\textbf {\bibinfo {volume} {11}},\ \bibinfo
  {pages} {449} (\bibinfo {year} {2016})}\BibitemShut {NoStop}%
\bibitem [{\citenamefont {Butenko}\ and\ \citenamefont
  {R\"o{\ss}ler}(2013)}]{butenko2013}%
  \BibitemOpen
  \bibfield  {author} {\bibinfo {author} {\bibfnamefont {A.~B.}\ \bibnamefont
  {Butenko}}\ and\ \bibinfo {author} {\bibfnamefont {U.~K.}\ \bibnamefont
  {R\"o{\ss}ler}},\ }\href@noop {} {\bibfield  {journal} {\bibinfo  {journal}
  {EPJ Web of Conferences}\ }\textbf {\bibinfo {volume} {40}},\ \bibinfo
  {pages} {08006} (\bibinfo {year} {2013})}\BibitemShut {NoStop}%
\bibitem [{\citenamefont {Michels}\ \emph {et~al.}(2016)\citenamefont
  {Michels}, \citenamefont {Mettus}, \citenamefont {Honecker},\ and\
  \citenamefont {Metlov}}]{Michels2016PRB}%
  \BibitemOpen
  \bibfield  {author} {\bibinfo {author} {\bibfnamefont {A.}~\bibnamefont
  {Michels}}, \bibinfo {author} {\bibfnamefont {D.}~\bibnamefont {Mettus}},
  \bibinfo {author} {\bibfnamefont {D.}~\bibnamefont {Honecker}},\ and\
  \bibinfo {author} {\bibfnamefont {K.~L.}\ \bibnamefont {Metlov}},\ }\href
  {https://doi.org/10.1103/PhysRevB.94.054424} {\bibfield  {journal} {\bibinfo
  {journal} {Phys. Rev. B}\ }\textbf {\bibinfo {volume} {94}},\ \bibinfo
  {pages} {054424} (\bibinfo {year} {2016})}\BibitemShut {NoStop}%
\bibitem [{\citenamefont {Michels}\ \emph {et~al.}(2019)\citenamefont
  {Michels}, \citenamefont {Mettus}, \citenamefont {Titov}, \citenamefont
  {Malyeyev}, \citenamefont {Bersweiler}, \citenamefont {Bender}, \citenamefont
  {Peral}, \citenamefont {Birringer}, \citenamefont {Quan}, \citenamefont
  {Hautle}, \citenamefont {Kohlbrecher}, \citenamefont {Honecker},
  \citenamefont {Fern\'andez}, \citenamefont {Barqu\'{\i}n},\ and\
  \citenamefont {Metlov}}]{Michels2019PRB}%
  \BibitemOpen
  \bibfield  {author} {\bibinfo {author} {\bibfnamefont {A.}~\bibnamefont
  {Michels}}, \bibinfo {author} {\bibfnamefont {D.}~\bibnamefont {Mettus}},
  \bibinfo {author} {\bibfnamefont {I.}~\bibnamefont {Titov}}, \bibinfo
  {author} {\bibfnamefont {A.}~\bibnamefont {Malyeyev}}, \bibinfo {author}
  {\bibfnamefont {M.}~\bibnamefont {Bersweiler}}, \bibinfo {author}
  {\bibfnamefont {P.}~\bibnamefont {Bender}}, \bibinfo {author} {\bibfnamefont
  {I.}~\bibnamefont {Peral}}, \bibinfo {author} {\bibfnamefont
  {R.}~\bibnamefont {Birringer}}, \bibinfo {author} {\bibfnamefont
  {Y.}~\bibnamefont {Quan}}, \bibinfo {author} {\bibfnamefont {P.}~\bibnamefont
  {Hautle}}, \bibinfo {author} {\bibfnamefont {J.}~\bibnamefont {Kohlbrecher}},
  \bibinfo {author} {\bibfnamefont {D.}~\bibnamefont {Honecker}}, \bibinfo
  {author} {\bibfnamefont {J.~R.}\ \bibnamefont {Fern\'andez}}, \bibinfo
  {author} {\bibfnamefont {L.~F.}\ \bibnamefont {Barqu\'{\i}n}},\ and\ \bibinfo
  {author} {\bibfnamefont {K.~L.}\ \bibnamefont {Metlov}},\ }\href
  {https://doi.org/10.1103/PhysRevB.99.014416} {\bibfield  {journal} {\bibinfo
  {journal} {Phys. Rev. B}\ }\textbf {\bibinfo {volume} {99}},\ \bibinfo
  {pages} {014416} (\bibinfo {year} {2019})}\BibitemShut {NoStop}%
\bibitem [{\citenamefont {Vedmedenko}\ \emph {et~al.}(2019)\citenamefont
  {Vedmedenko}, \citenamefont {Riego}, \citenamefont {Arregi},\ and\
  \citenamefont {Berger}}]{Vedmedenko2019PRL}%
  \BibitemOpen
  \bibfield  {author} {\bibinfo {author} {\bibfnamefont {E.~Y.}\ \bibnamefont
  {Vedmedenko}}, \bibinfo {author} {\bibfnamefont {P.}~\bibnamefont {Riego}},
  \bibinfo {author} {\bibfnamefont {J.~A.}\ \bibnamefont {Arregi}},\ and\
  \bibinfo {author} {\bibfnamefont {A.}~\bibnamefont {Berger}},\ }\href
  {https://doi.org/10.1103/PhysRevLett.122.257202} {\bibfield  {journal}
  {\bibinfo  {journal} {Phys. Rev. Lett.}\ }\textbf {\bibinfo {volume} {122}},\
  \bibinfo {pages} {257202} (\bibinfo {year} {2019})}\BibitemShut {NoStop}%
\bibitem [{\citenamefont {Avci}\ \emph {et~al.}(2021)\citenamefont {Avci},
  \citenamefont {Lambert}, \citenamefont {Sala},\ and\ \citenamefont
  {Gambardella}}]{Avci2021}%
  \BibitemOpen
  \bibfield  {author} {\bibinfo {author} {\bibfnamefont {C.~O.}\ \bibnamefont
  {Avci}}, \bibinfo {author} {\bibfnamefont {C.-H.}\ \bibnamefont {Lambert}},
  \bibinfo {author} {\bibfnamefont {G.}~\bibnamefont {Sala}},\ and\ \bibinfo
  {author} {\bibfnamefont {P.}~\bibnamefont {Gambardella}},\ }\href
  {https://doi.org/10.1103/PhysRevLett.127.167202} {\bibfield  {journal}
  {\bibinfo  {journal} {Phys. Rev. Lett.}\ }\textbf {\bibinfo {volume} {127}},\
  \bibinfo {pages} {167202} (\bibinfo {year} {2021})}\BibitemShut {NoStop}%
\bibitem [{\citenamefont {Michels}\ \emph {et~al.}(2014)\citenamefont
  {Michels}, \citenamefont {Erokhin}, \citenamefont {Berkov},\ and\
  \citenamefont {Gorn}}]{michels2014jmmm}%
  \BibitemOpen
  \bibfield  {author} {\bibinfo {author} {\bibfnamefont {A.}~\bibnamefont
  {Michels}}, \bibinfo {author} {\bibfnamefont {S.}~\bibnamefont {Erokhin}},
  \bibinfo {author} {\bibfnamefont {D.}~\bibnamefont {Berkov}},\ and\ \bibinfo
  {author} {\bibfnamefont {N.}~\bibnamefont {Gorn}},\ }\href@noop {} {\bibfield
   {journal} {\bibinfo  {journal} {J. Magn. Magn. Mater.}\ }\textbf {\bibinfo
  {volume} {350}},\ \bibinfo {pages} {55} (\bibinfo {year} {2014})}\BibitemShut
  {NoStop}%
\bibitem [{\citenamefont {Maleev}(2002)}]{maleyev2002}%
  \BibitemOpen
  \bibfield  {author} {\bibinfo {author} {\bibfnamefont {S.~V.}\ \bibnamefont
  {Maleev}},\ }\href@noop {} {\bibfield  {journal} {\bibinfo  {journal}
  {Physics--Uspekhi}\ }\textbf {\bibinfo {volume} {45}},\ \bibinfo {pages}
  {569} (\bibinfo {year} {2002})}\BibitemShut {NoStop}%
\bibitem [{\citenamefont {Fedorov}\ \emph {et~al.}(1997)\citenamefont
  {Fedorov}, \citenamefont {Gukasov}, \citenamefont {Kozlov}, \citenamefont
  {Maleyev}, \citenamefont {Plakhty},\ and\ \citenamefont
  {Zobkalo}}]{Fedorov1997}%
  \BibitemOpen
  \bibfield  {author} {\bibinfo {author} {\bibfnamefont {V.~I.}\ \bibnamefont
  {Fedorov}}, \bibinfo {author} {\bibfnamefont {A.~G.}\ \bibnamefont
  {Gukasov}}, \bibinfo {author} {\bibfnamefont {V.}~\bibnamefont {Kozlov}},
  \bibinfo {author} {\bibfnamefont {S.~V.}\ \bibnamefont {Maleyev}}, \bibinfo
  {author} {\bibfnamefont {V.~P.}\ \bibnamefont {Plakhty}},\ and\ \bibinfo
  {author} {\bibfnamefont {I.~A.}\ \bibnamefont {Zobkalo}},\ }\href@noop {}
  {\bibfield  {journal} {\bibinfo  {journal} {Phys. Lett. A}\ }\textbf
  {\bibinfo {volume} {224}},\ \bibinfo {pages} {372} (\bibinfo {year}
  {1997})}\BibitemShut {NoStop}%
\bibitem [{\citenamefont {Grigoriev}\ \emph {et~al.}(2008)\citenamefont
  {Grigoriev}, \citenamefont {Chetverikov}, \citenamefont {Lott},\ and\
  \citenamefont {Schreyer}}]{lott08}%
  \BibitemOpen
  \bibfield  {author} {\bibinfo {author} {\bibfnamefont {S.~V.}\ \bibnamefont
  {Grigoriev}}, \bibinfo {author} {\bibfnamefont {Y.~O.}\ \bibnamefont
  {Chetverikov}}, \bibinfo {author} {\bibfnamefont {D.}~\bibnamefont {Lott}},\
  and\ \bibinfo {author} {\bibfnamefont {A.}~\bibnamefont {Schreyer}},\
  }\href@noop {} {\bibfield  {journal} {\bibinfo  {journal} {Phys. Rev. Lett.}\
  }\textbf {\bibinfo {volume} {100}},\ \bibinfo {pages} {197203} (\bibinfo
  {year} {2008})}\BibitemShut {NoStop}%
\bibitem [{\citenamefont {Legvold}(1980)}]{legvold80}%
  \BibitemOpen
  \bibfield  {author} {\bibinfo {author} {\bibfnamefont {S.}~\bibnamefont
  {Legvold}},\ }in\ \href@noop {} {\emph {\bibinfo {booktitle} {Handbook of
  Magnetic Materials}}},\ Vol.~\bibinfo {volume} {1},\ \bibinfo {editor}
  {edited by\ \bibinfo {editor} {\bibfnamefont {E.~P.}\ \bibnamefont
  {Wohlfarth}}}\ (\bibinfo  {publisher} {North-Holland Publishing Company},\
  \bibinfo {address} {Amsterdam},\ \bibinfo {year} {1980})\ pp.\ \bibinfo
  {pages} {183--295}\BibitemShut {NoStop}%
\bibitem [{\citenamefont {Chikazumi}(1997)}]{chikazumi1997book}%
  \BibitemOpen
  \bibfield  {author} {\bibinfo {author} {\bibfnamefont {S.}~\bibnamefont
  {Chikazumi}},\ }\href@noop {} {\emph {\bibinfo {title} {{Physics of
  Ferromagnetism}}}}\ (\bibinfo  {publisher} {Oxford University Press},\
  \bibinfo {address} {Oxford},\ \bibinfo {year} {1997})\BibitemShut {NoStop}%
\bibitem [{\citenamefont {Hegland}\ \emph {et~al.}(1963)\citenamefont
  {Hegland}, \citenamefont {Legvold},\ and\ \citenamefont
  {Spedding}}]{hegland63}%
  \BibitemOpen
  \bibfield  {author} {\bibinfo {author} {\bibfnamefont {D.~E.}\ \bibnamefont
  {Hegland}}, \bibinfo {author} {\bibfnamefont {S.}~\bibnamefont {Legvold}},\
  and\ \bibinfo {author} {\bibfnamefont {F.~H.}\ \bibnamefont {Spedding}},\
  }\href@noop {} {\bibfield  {journal} {\bibinfo  {journal} {Phys. Rev.}\
  }\textbf {\bibinfo {volume} {131}},\ \bibinfo {pages} {158} (\bibinfo {year}
  {1963})}\BibitemShut {NoStop}%
\bibitem [{\citenamefont {Garanin}(1996)}]{garanin1992}%
  \BibitemOpen
  \bibfield  {author} {\bibinfo {author} {\bibfnamefont {D.~A.}\ \bibnamefont
  {Garanin}},\ }\href {https://doi.org/10.1103/PhysRevB.53.11593} {\bibfield
  {journal} {\bibinfo  {journal} {Phys. Rev. B}\ }\textbf {\bibinfo {volume}
  {53}},\ \bibinfo {pages} {11593} (\bibinfo {year} {1996})}\BibitemShut
  {NoStop}%
\bibitem [{\citenamefont {Evans}\ \emph {et~al.}(2014)\citenamefont {Evans},
  \citenamefont {Fan}, \citenamefont {Chureemart}, \citenamefont {Ostler},
  \citenamefont {Ellis},\ and\ \citenamefont {Chantrell}}]{evans2014}%
  \BibitemOpen
  \bibfield  {author} {\bibinfo {author} {\bibfnamefont {R.~F.~L.}\
  \bibnamefont {Evans}}, \bibinfo {author} {\bibfnamefont {W.~J.}\ \bibnamefont
  {Fan}}, \bibinfo {author} {\bibfnamefont {P.}~\bibnamefont {Chureemart}},
  \bibinfo {author} {\bibfnamefont {T.~A.}\ \bibnamefont {Ostler}}, \bibinfo
  {author} {\bibfnamefont {M.~O.~A.}\ \bibnamefont {Ellis}},\ and\ \bibinfo
  {author} {\bibfnamefont {R.~W.}\ \bibnamefont {Chantrell}},\ }\href
  {https://doi.org/10.1088/0953-8984/26/10/103202} {\bibfield  {journal}
  {\bibinfo  {journal} {J. Phys.: Condens. Matter}\ }\textbf {\bibinfo {volume}
  {26}},\ \bibinfo {pages} {103202} (\bibinfo {year} {2014})}\BibitemShut
  {NoStop}%
\end{thebibliography}

%

\end{document}